\documentclass{ws-sd}
 \usepackage{epsfig,subfigure,color,amsmath}

\begin{document}

\markboth{P. Ditlevsen and H. Braun}{Pseudo resonance induced quasi-periodic behavior in stochastic threshold dynamics}

\catchline{}{}{}{}{}

\title{PSEUDO RESONANCE INDUCED QUASI-PERIODIC BEHAVIOR IN STOCHASTIC THRESHOLD DYNAMICS}

\author{PETER D. DITLEVSEN and HOLGER BRAUN}

\address{Centre for Ice and Climate, The Niels
Bohr Institute,\\ University of Copenhagen, Juliane Maries Vej 30,\\
 DK-2100 Copenhagen O, Denmark\\pditlev@nbi.ku.dk}



\maketitle

\begin{history}
\received{(Day Month Year)}
\revised{(Day Month Year)}
\end{history}

\begin{abstract}
Here we present a simple stochastic threshold model consisting of a deterministic
slowly decaying term and a fast stochastic noise term. The process shows a
pseudo-resonance, in the sense that for small and large intensities of the noise 
the signal is irregular and the distribution of threshold crossings is broad, while
for a tuned intermediate value of noise intensity the signal becomes quasi-periodic and 
the distribution of threshold crossings is narrow. The mechanism captured by the model might be relevant for explaining apparent quasi-periodicity of observed climatic
variations where no internal or external  periodicities can be identified. 
\end{abstract}


\ccode{}

\section{Introduction}	
In many cases the dynamics of complex systems are only partially understood from governing principles.
Often the dynamics has to be inferred from observed behavior of the system. Typically a time series of
measurements or observations of a representative parameter must be analyzed. 
Cyclic variations are often observed in natural systems; linear harmonic oscillators, non-linear oscillators and limit cycles
are internal to the system while phenomena like stochastic resonance (SR)\cite{benzi:1982} is dynamical amplification of a
weak external periodic forcing. Another situation is coherence resonance\cite{pikovski:1997} where a limit cycle in the Fitz Hugh-Nagumo system is excited by a stochastic noise. 
Identification of periodic behavior is appealing because it provides a possibility of predicting future development of the system, however, for noisy systems observed over a limited time period competing models might be equally well matching the observations. One example is the famous Canadian Lynx catch time series: This has been explained by variations of Lotka-Voltarra non-linear oscillator phenomenon \cite{ayala:1973}  and alternatively as a result of a threshold autoregressive (TAR) process \cite{tong:1980}. Likewise, the apparent regularity of climate variations over time cannot presently be
reproduced from first principles. Thus we rely on identifying mechanisms governing the behavior from 
simpler models. 

\section{Climate cycles}
The climate has changed in the ice age cycles with a characteristic time scale of approximately 100 kyr (kilo-year) over the past million years. These
variations are attributed to changes in the incoming solar radiation, the insolation, resulting from changes in Earth's orbit around the sun. The main variations are due to the precession of the
axis of rotation with periods around 20 kyr, the harmonically changing inclination of the 
axis of rotation with respect to the ecliptic plane, the obliquity, with a period of 41 kyr and finally
the changes in the eccentricity of the orbit, which has several harmonic components, all in the order
of 100 kyr. The variation of ice ages is of the order of the eccentricity cycle. This has long been
a mystery, since the magnitude of variations in insolation due to this cycle is an order of magnitude
smaller than that of the other two orbital cycles. 

The mechanism of stochastic resonance (SR) \cite{benzi:1982} was proposed to solve this enigma. SR is a mechanism for noise assisted amplifying a non-linear response in a system to a (weak) periodic forcing.
SR does not offer an explanation for why the dominant 20 kyr and 41 kyr
periods do not dominate the late Pleistocene ice age cycles. Today, this is not considered
as a plausible explanation for the ice ages and 
it is generally accepted that the 100 kyr glacial time scale cannot be attributed to 
the eccentricity cycle \cite{huybers:2007}. 
In the Plio- and early 
Pleistocene, 3--1 My BP, the dominant period of variation was indeed the 41 kyr 
obliquity variation. 

The glacial cycles have been proposed to be a result of the glacial ice volume being a TAR process 
\cite{huybers:2005}
with a constant drift until a threshold is reached, where the ice sheets collapse and the 
process is reset to zero. In the low noise limit, this process becomes periodic, with the
period set by the ratio of the threshold (maximum ice volume) and the constant
drift (ice growth). Here, in contrast to the SR model, an external
periodic driver need not to be assumed. The Achilles heal of this model is the interpretation 
in terms of ice sheet build up
and collapse. The natural time scale for build up of the big ice sheets is an order of 
magnitude shorter than 100 kyr.  

Within the last glacial period, which is recorded in high resolution paleoclimatic ice cores and ocean sediments, more or less regular episodes of warmer climates, the Dansgaards-Oeschger (DO) events \cite{dansgaard:1993}, are seen.  These occur at the millennial time scale more or less regularly.  A quasi-periodicity of 1470 years has been proposed \cite{schulz:2002}, 
this is, however, not significantly different from what would be expected from a purely
random occurrence \cite{ditlevsen:2007}. For the DO events no external periodic forcing has 
been identified, though a ghost resonance as a response to the beating of periodic solar luminosity variations has been proposed \cite{braun:2005}.

Here we will propose a stochastic threshold model, which shows a pseudo resonance behavior in such a way that the waiting times between threshold crossings are irregular for small and large noise intensities, while the signal becomes quasi-periodic for intermediate values of the noise intensity. This kind of apparent periodicity arising with no external periodic forcing could explain the observed apparent regularity in the climate record. It is distinct from a stochastic resonance phenomenon where an external periodic forcing is amplified.

\section{The quasi-periodic stochastic dynamics}
We propose the following stochastic threshold process,

\begin{eqnarray}
 dx&= &\left\{\begin{array}{ll}
-(x-x_1)dt/\tau, & x+y< x_2 \\
-x_2,& x+y\ge x_2 \end{array} \right. \\ 
dy&=&-\alpha y dt + \tilde{\sigma} dB
\end{eqnarray}
with $(x(0),y(0))=(0,0)$. The "fast" process $y(t)$ is a simple Ohrnstein-Uhlenbeck process with variance $\sigma^2=\tilde{\sigma}^2/2\alpha$, $\alpha^{-1} \ll \tau$. $dB$ is the 
usual Brownian white noise.
For the "slow" process $x(t)$ the scalar $x_2$ is a threshold with $x_2>x_1>0$. The model can as well
be seen as consisting of a purely deterministic relaxation process $x$ and a stochastic threshold
$x_2-y$.    

In the following we shall compare the stochastic threshold model with the 
thres\-hold autoregressive (TAR) model. The TAR model is defined as the following discrete stochastic process:

\begin{equation}
x_n=\left\{\begin{array}{ll} 
a_0+a_1x_{n-1}+ ... + a_mx_{n-m}+\sigma_1\eta_n, & x_{n-1} < z \\
           b_0+b_1x_{n-1}+ ... + b_mx_{n-m}+\sigma_2\eta_n,   & x_{n-1} \ge  z\end{array} \right. \\
\end{equation}
where $\eta_n$ is a unit variance uncorrelated gaussian noise, $z$ is a threshold and the $a$'s and $b$'s
are linear regression coefficients. Here we shall only consider the simple case of a linear trend TAR
\cite{huybers:2005}: 

\begin{eqnarray}
x_n=\left\{\begin{array}{ll}
a+x_{n-1}+\sigma \eta_n, & x_{n-1} < z \\
         0, &  x_{n-1} \ge  z \end{array}\right. \\
\end{eqnarray}
This process has, in the limit of small noise $\sigma \rightarrow 0$, a saw-tooth shape with
period $T=z/a$. 
The times $T_c$ where the threshold is crossed is another stochastic variable, the distance between which we denote "waiting
times". At each threshold crossing we may reset time $n\rightarrow n-T_c$. 
We shall be interested in the distribution of the waiting times. This is equivalent to a 
first passage problem. These are notoriously difficult to solve analytically, but easy
to simulate: In figure 1, left panels, realizations of the stochastic threshold process are shown for increasing values of the variance $\sigma$.
For visibility the processes are shown in light grey, while smoothed signals are shown in black.  
These are compared to realizations of the linear drift TAR process, right panels. It is seen that for 
the intermediate value of the variance the threshold crossing is most periodic. This is in contrast to 
the case of the linear drift TAR process in the right panels. In this case the signal becomes perfectly
periodic in the limit $\sigma=0$. Figure 2 shows the waiting time distributions for for the six cases calculated
from simulations (much longer than the ones shown).   
 
In the limit $\sigma = 0$ the process $x$ will approach $x_1$: 
$x(t)=x_1(1-\exp(-t/\tau))$ and never reach the 
threshold $x_2$. 
In the small noise limit the process will reach the asymptote $x(t)\rightarrow x_1$
before a noise induced crossing of the level $x_2$. We can then estimate the mean waiting time for crossing  from the stationary distribution of $y(t)$: 

$$p(\sigma)\equiv \text{Prob}(y>x_2-x_1)=\Phi((x_2-x_1)/\sigma)=\frac{1}{\sqrt{2\pi}}\int_{-\infty}^{(x_2-x_1)/\sigma}
e^{-x^2/2}dx$$ 
where $\Phi(x)$ is the error-function. The correlation time for $y(t)$ is $\tau_y=
\alpha^{-1}$, we can thus estimate the mean waiting time $\langle T\rangle$ from a discrete series, where the 
(independent) probability of up-crossing in any time interval $n \tau_y < t \le (n+1)\tau_y$ is $p(\sigma)$:

$$ \langle T\rangle\approx \tau_y \sum_{n=1}^\infty n (1-p(\sigma))^{n-1}*p(\sigma)=\frac{\tau_y}{1-p(\sigma)}$$
Similarly in the large noise limit, the mean waiting time is determined by the waiting time for $y(t)$ to
exceed $x_2$, thus we get
$$\langle T\rangle\approx \frac{\tau_y}{1-q(\sigma)}$$
where $q(\sigma)=\Phi(x_2/\sigma)$. The two estimates are shown in figure 3, top left panel, where the
thick curve is the mean waiting time for the process obtained by simulation. 
The dashed line corresponds to the natural time scale of the decay process $\tau$. This time scale does not 
govern the pseudo-resonance. The same phenomenon is seen for a deterministic process $\tilde{x}(t)$, such as $\tilde{x}(t)=x_1(1-(t+1)^{-\gamma})$, with $\gamma>0$, where no natural time scale can be defined.
 
In the case of the linear drift
TAR process the mean waiting time is independent of the noise level (figure 3, top right panel). We define
the variance of the waiting time distributions (shown in figure 2) to quantify the periodicity of the signals. 
The bottom left panel in figure 3 shows the this for the stochastic threshold model as a function of the variance $\sigma$. It is seen that there is indeed a distinct minimum, or pseudo resonance, corresponding to a
quasi-periodicity as a function of the noise intensity. The corresponding quantity for the linear trend TAR is shown in the bottom right panel, showing that the signal becomes periodic as the intensity of the noise becomes small.

In the case of the climatic  DO events some climate models suggest that
the onset
of the events can be regarded as a shift into a non-equilibrium state of the
climate system, and that the return to equilibrium happens in the form of
a relaxation process. This makes our simple model a plausible candidate for explaining the apparent regularity in the climate record. In figure 4 top panel a sequence of the climate record 
from the NGRIP ice core from Greenland\cite{ngrip:2004}  is shown. The $\delta^{18}O$ record is an
isotopic proxy for temperature, so the record shows rapid changes into a warm climate 
followed by slow relaxations toward the cold glacial climate. The bottom panel shows
for comparison
a realization of the stochastic threshold model with $\sigma=0.4$ and $\tau=3000$.  Note
that $-x$ is plotted in order to have the jump in the positive direction.

In summary,
we have introduced a stochastic threshold model, which exhibits a pseudo resonance quasi-periodicity as a function of the noise intensity. The model is not forced
at that frequency, and thus offers an explanation of apparent regularity in observed time series completely
different from that of the linear trend TAR model and the stochastic resonance model.

\newpage \begin{center} FIGURE CAPTIONS \end{center}
\newcounter{fig} \begin{list}{Fig. \arabic{fig}}
{\usecounter{fig}\setlength{\labelwidth}{2cm}\setlength{\labelsep}{3mm}}

\item
Left columns show
realizations of the stochastic threshold model for increasing values of the noise intensity
$\sigma$.
For small and large noise intensities the signal is irregular while for an optimally chosen 
intermediate value the signal is quasi-periodic. The signals are in light gray, for better 
visibility 10 point running mean curves are shown in black. The other parameters used in the simulations are
$x_1=1.5, x_2=2, \tau=100, \alpha=1$.
Right columns show similar
realizations for the linear trend TAR with $a=0.01$. In this case the signal becomes increasingly periodic
as the noise intensity decreases.  

\item 
The probability densities for the up-crossing waiting times in the six cases shown in figure 1.
These are calculated from much longer simulations that the ones shown in figure 1. 

\item 
Top left panel shows the mean waiting time for the stochastic threshold model as a function
 of noise intensity $\sigma$. The thin curves are the estimates for the low (left) and high (right) limits of noise intensities. Bottom left panel shows the width of the waiting time distributions (shown for three values of $\sigma$ in figure 2). The minimum defines the
resonance with a quasi-periodic signal (middle left panel in figures 1 and 2). The
right panels show the similar graphs for the linear trend TAR. In this case the $\sigma=0$
is perfectly periodic.

\item
Top panel shows a part of the climate record 
from the NGRIP ice core from Greenland\cite{ngrip:2004} . The $\delta^{18}O$ record is an
isotopic proxy for temperature, so the record shows rapid changes into a warm climate 
followed by slow relaxations toward the cold glacial climate. The bottom panel shows
a smoothed realization of the stochastic threshold model with $\sigma=0.4$ and $\tau=3000$.  Note that $-x$ is plotted.

\end{list}

\newpage 
\begin{figure}[!H] \begin{center}\epsfxsize=8cm 
\epsffile{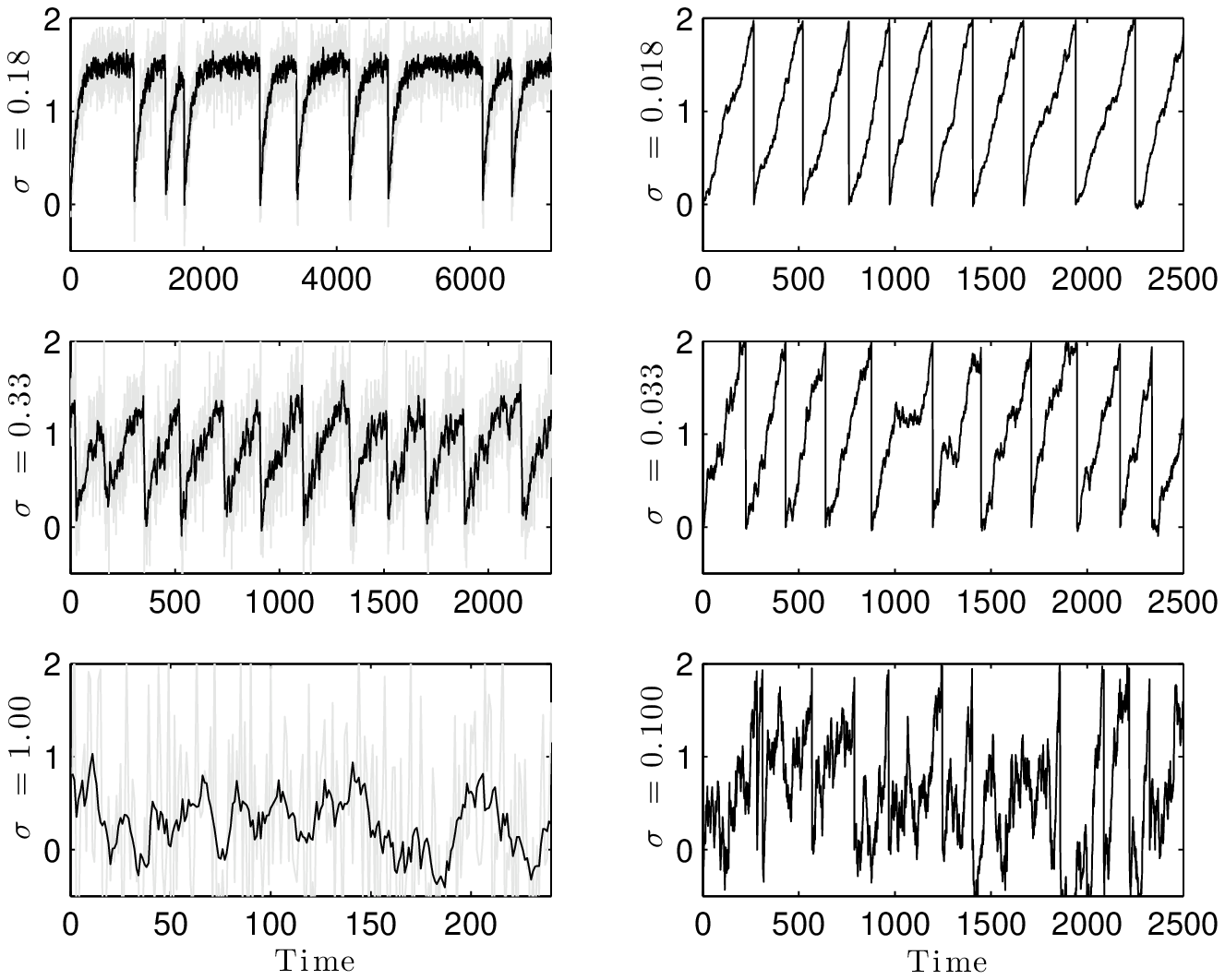}\caption[]{
}
\label{fig1} \end{center}\end{figure}

\begin{figure}[!H] \begin{center}\epsfxsize=8cm 
\epsffile{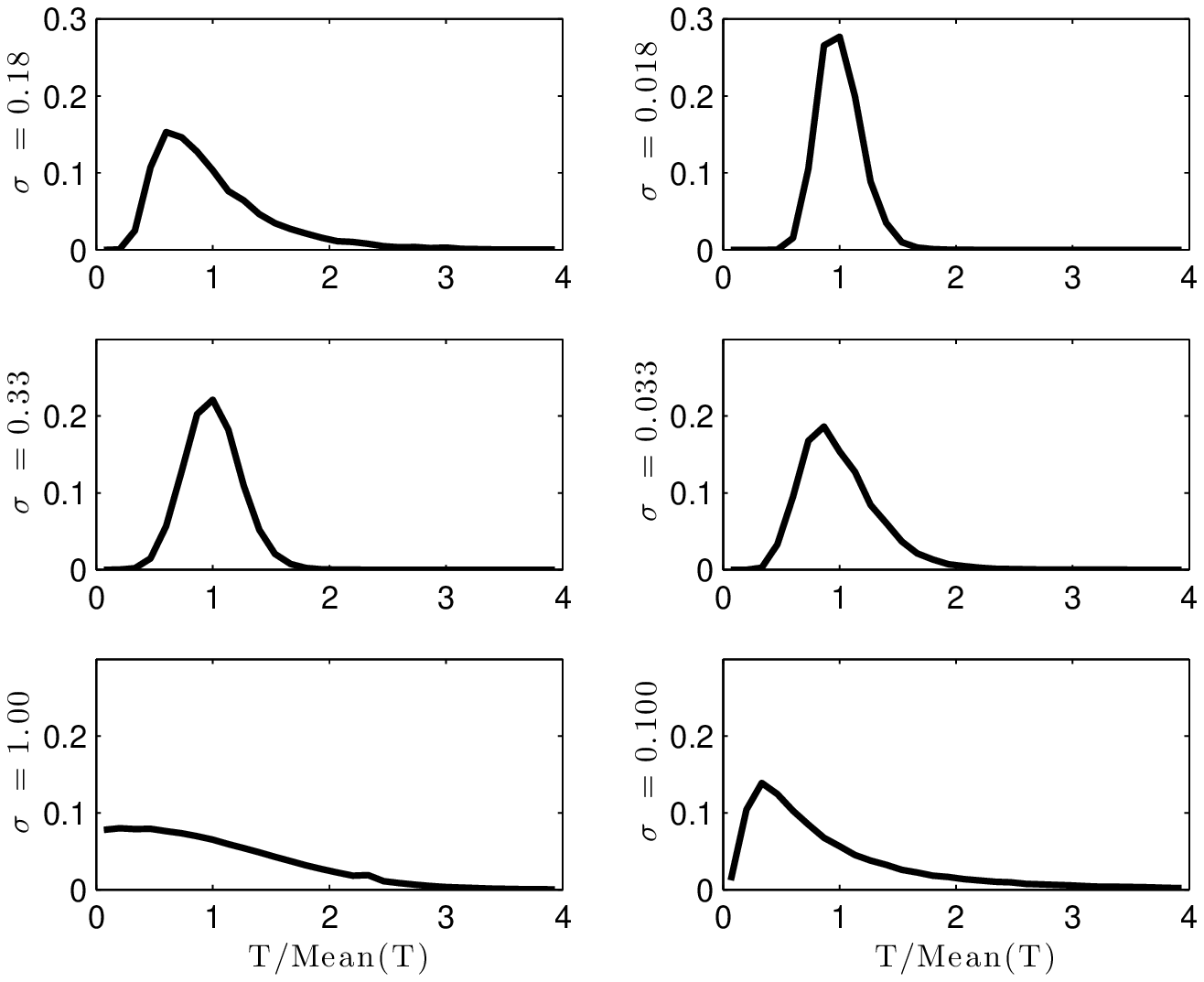}\caption[]{
}
\label{fig2} \end{center}\end{figure}

\begin{figure}[!H] \begin{center}\epsfxsize=8cm 
\epsffile{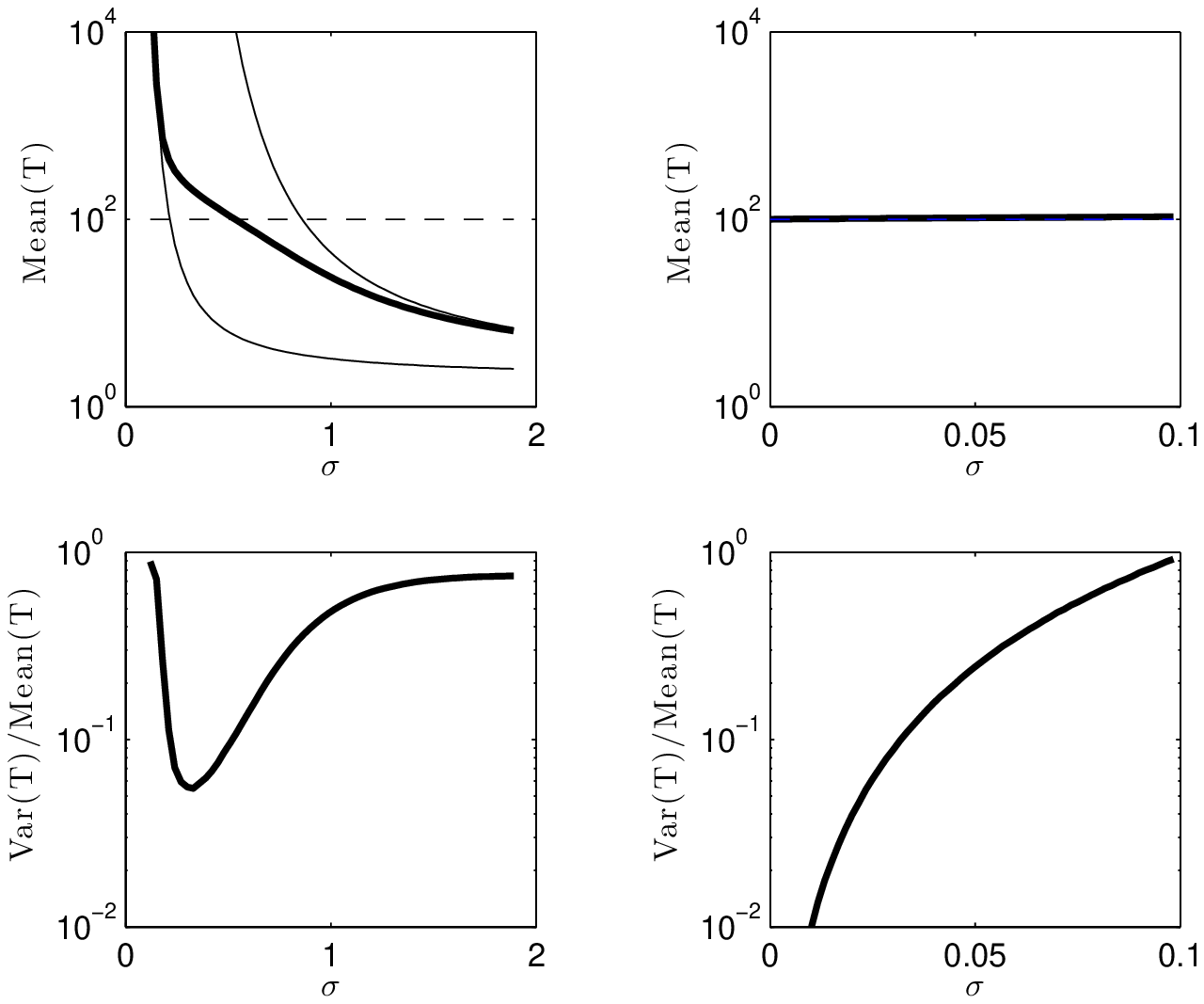}\caption[]{
}
\label{fig3} \end{center}\end{figure}

\begin{figure}[!H] \begin{center}\epsfxsize=8cm 
\epsffile{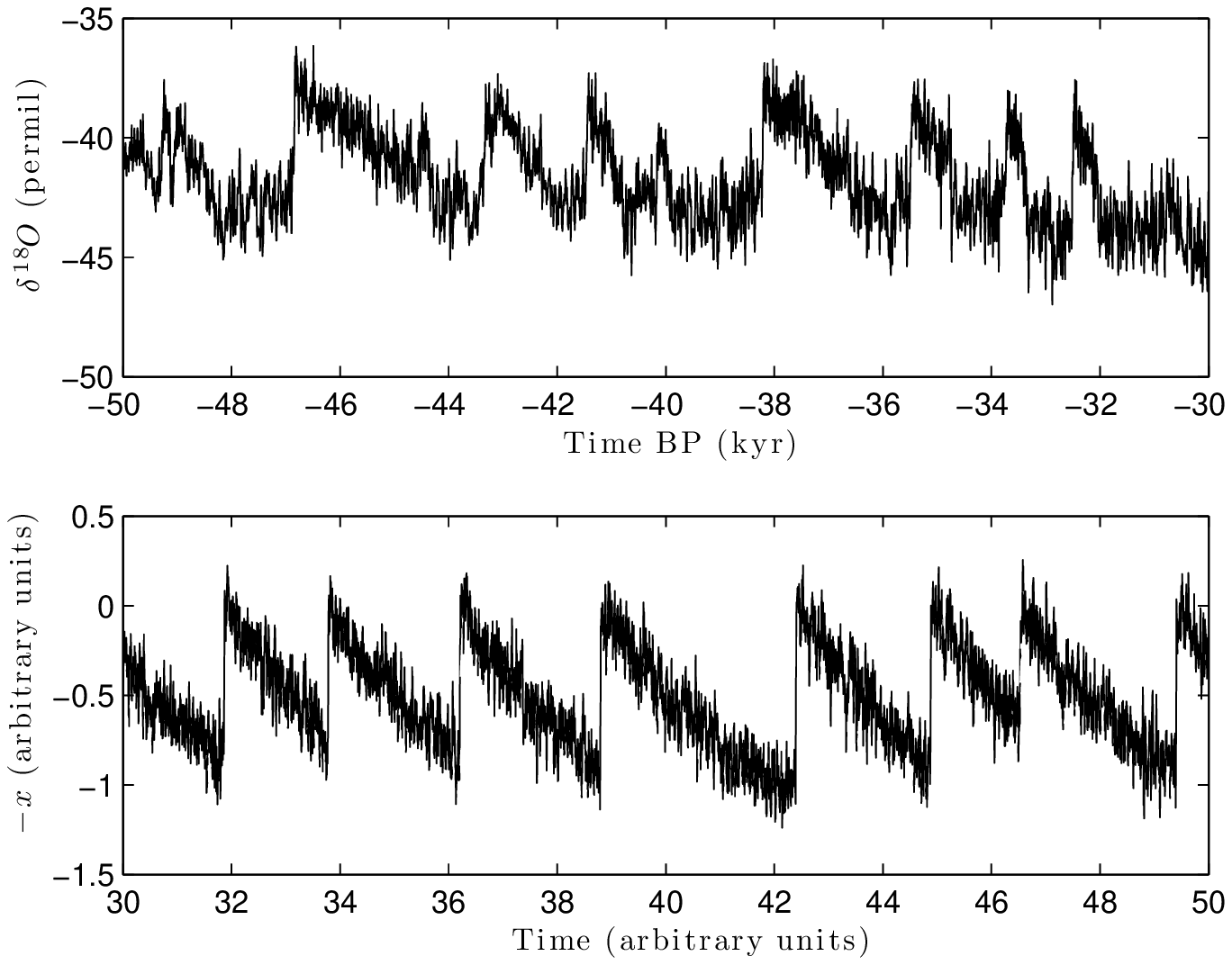}\caption[]{
}
\label{fig4} \end{center}\end{figure}

\end{document}